\documentstyle{article}
\font\tenbf=cmbx10
\font\tenrm=cmr10
\font\tenit=cmti10
\font\elevenbf=cmbx10 scaled\magstep 1
\font\elevenrm=cmr10 scaled\magstep 1
\font\elevenit=cmti10 scaled\magstep 1

\renewenvironment{thebibliography}[1]
 { \elevenrm
   \begin{list}{\arabic{enumi}.}
    {\usecounter{enumi} \setlength{\parsep}{0pt}
     \setlength{\itemsep}{3pt} \settowidth{\labelwidth}{#1.}
     \sloppy
    }}{\end{list}}

\newcommand{\Rn}[1]{{\uppercase\expandafter{\romannumeral#1}}}
\newcommand{\gsim}%
{\mbox{\raisebox{-1.0ex}
    {$\ \stackrel{\textstyle >}{\textstyle \sim}\ $}}}
\newcommand{\lsim}%
 {\mbox{\raisebox{-1.0ex}
     {$\ \stackrel{\textstyle <}{\textstyle \sim}\ $}}}

\newcommand{\etal}{{\elevenit et al.\/}}

\newcommand{\bsg}{$b\rightarrow s\ \gamma$\ }

\newcommand{\Journal}[4]{{#1} {\elevenbf #2} {(#3)} {#4}}

\newcommand{\plb}{\elevenit Phys.~Lett.~{\elevenbf B}}
\newcommand{\prp}{\elevenit Phys.~Rep.}

\newcommand{\prd}{\elevenit Phys.~Rev.~{\elevenbf D}}
\newcommand{\prl}{\elevenit Phys.~Rev.~Lett.}

\newcommand{\npb}{\elevenit Nucl.~Phys.~{\elevenbf B}}
\newcommand{\ptp}{\elevenit Prog.~Theor.~Phys.}

\newcommand{\zpc}{\elevenit Z.~Phys.~{\elevenbf C}}

\catcode`\@=11
\def\makepreprititle{\par
  \begingroup
  \def\thefootnote{\fnsymbol{footnote}}
  \def\
@makefnmark{\hbox
  to 0pt{$^{\@thefnmark}$\hss}}
  \if@twocolumn
  \twocolumn[\@makepreprititle]
  \else \newpage
  \global\@topnum\z@
  \@makepreprititle \fi\thispagestyle{empty}\@thanks
  \endgroup
  \setcounter{footnote}{0}
  \let\makepreprititle\relax
  \let\@makepreprititle\relax
  \gdef\@thanks{}\gdef\@author{}\gdef\@title{}
  \gdef\@preprintnumber{}\gdef\@preprintdate{}\gdef\subtitle{}
  \let\thanks\relax}
\def\preprintnumber#1{\gdef\@preprintnumber{#1}}
\def\preprintdate#1{\gdef\@preprintdate{#1}}
\def\subtitle#1{\gdef\@subtitle{#1}}
\def\@makepreprititle{\newpage
{\def\baselinestretch{1}
  \begin{flushright} \@preprintnumber \par
  \@preprintdate \end{flushright} } \par
  \begin{center}
\vskip 1.5em
  {\LARGE \@title \par} \vskip 2.5em
  {\Large \lineskip .5em
  \begin{tabular}[t]{c}\@author
  \end{tabular}\par}
  \vskip 1em {\large \@date} \end{center}
  \par
  \vfil}
\date{\sl Theory Group, KEK, Tsukuba, Ibaraki, 305, Japan}
\preprintdate{~}
\preprintnumber{~}
\subtitle{}
\def\abstract{\if@twocolumn
\section*{Abstract}
\else \normalsize
\begin{center}
{\bf Abstract\vspace{-.5em}\vspace{0pt}}
\end{center}
\quotation
\addtocounter{page}{-1}
\fi}
\def\endabstract{\if@twocolumn\else\endquotation\fi}
\def\spacing#1{\def\baselinestretch{#1}
\typeout{baselinestretch is modified to \baselinestretch}}
\catcode`\@=12
\spacing{1.4}
\preprintnumber{KEK-TH-428\\
                KEK Preprint 94-192}
\preprintdate{February, 1995}
\title{ SUSY and CP Violation\footnote{Talk given at 1994
International Workshop on B Physics, October 26-28 1994,
Nagoya, Japan.}}
\author{Yasuhiro~Okada}
\begin{document}
\makepreprititle
\begin{abstract}
Flavor changing neutral current and CP violating processes are
discussed in the minimal supergravity model. The constraint on
charged Higgs mass from the new measurement of the inclusive
branching ratio of the $b\rightarrow s\gamma$ process is obtained.
The $B^0_d-\bar{B}^0_d$ mixing parameter ($x_d$)
and the CP violating parameter in the $K^0-\bar{K}^0$ mixing
($\epsilon _K$) are calculated in this model and it is shown
that these parameters can be enhanced by 10\% $\sim$ 20 \%
compared to the prediction within the standard model.
Impacts on new physics search at B factories are also discussed.
\par
\vfil
\end{abstract}
\newpage
\begin{center}{{\tenbf SUSY and CP Violation\\}
\vglue 5pt
\vglue 1.0cm
{\tenrm YASUHIRO OKADA \\}
\baselineskip=13pt
{\tenit Theory Group, KEK, Tsukuba, Ibaraki, 305, Japan\\}
\baselineskip=12pt
\vglue 0.8cm
{\tenrm ABSTRACT}}
\end{center}
\vglue 0.3cm
{\rightskip=3pc
 \leftskip=3pc
 \tenrm\baselineskip=12pt
 \noindent
Flavor changing neutral current and CP violating processes are
discussed in the minimal supergravity model. The constraint on
charged Higgs mass from the new measurement of the inclusive
branching ratio of the $b\rightarrow s\gamma$ process is obtained.
The $B^0_d-\bar{B}^0_d$ mixing parameter ($x_d$)
and the CP violating parameter in the $K^0-\bar{K}^0$ mixing
($\epsilon _K$) are calculated in this model and it is shown
that these parameters can be enhanced by 10\% $\sim$ 20 \%
compared to the prediction within the standard model.
Impacts on new physics search at B factories are also discussed.

\vglue 0.6cm}
{\elevenbf\noindent 1. Introduction}
\vglue 0.4cm
\baselineskip=14pt
\elevenrm
Although nature of gauge interactions between fermions and gauge bosons has
been getting clearer in recent years, other aspects of the standard
model (SM) such as
origin of CP violation and mechanism of electroweak symmetry breaking have
not yet been understood experimentally. Main purpose of B factory projects at
KEK and SLAC is to measure CP violating asymmetries in B decays and
to see whether various CP violating processes are consistently explained
by a single phase of the Cabbibo-Kobayashi-Maskawa (CKM) matrix. If that
turns out to be the case, one of important aspects of the SM will be
established experimentally, on the other hand, if not, we will have a clear
evidence of physics beyond the SM.

One of promising candidates of the physics beyond the SM is a supersymmetric
(SUSY) extension. It is then important to know how the SUSY effects can
appear in physics on B decays. Although the original motivation of the SUSY
models is to give a possible explanation to the hierarchy problem in the
Higgs sector of the SM, their flavor sector also has unique
features. Firstly, any viable SUSY model contains at least two Higgs
doublets, therefore a physical charged Higgs is an inevitable consequence
of the SUSY SM.  This Higgs
can affect flavor changing neutral current (FCNC) processes through
loop diagrams. Moreover, squark mass matrices can be new sources of
flavor mixing. Since these squark mass matrices are in part determined from
SUSY
breaking parameters, the investigation of FCNC and CP violating processes could
lead us to knowledges about the SUSY breaking mechanism.

In this talk I will discuss how SUSY effects can appear in physics on
B decays. For this purpose, we should distinguish two scenarios on the
flavor mixing in SUSY models\cite{CPsusy}.
The first one is the `` minimal flavor
mixing case'' in which the origin of the flavor mixing and the CP
violation lies in the ordinary Yukawa coupling constants in superpotential.
The minimal supergravity model is classified as an example of this type,
where the SUSY breaking parameters are real and common for all flavors
at the grand unification scale. The other corresponds to the general flavor
mixing case where the flavor mixing in the squark sector is not simply
related to that in the quark sector and CP violations could arise from
many new physical complex phases. In the latter case  phenomenological
constraints from the $K^0-\bar{K}^0$ mixing and the neutron's electric
dipole moment  become important. Here we only consider the minimal case,
especially a model based on the minimal supergravity model. We first
consider the constraints on parameters obtained from the recent
measurement of the inclusive $b\rightarrow s\gamma$ branching ratio
\cite{cleo}.
Although there have been many works\cite{BBMR}-\cite{Carena}
on the SUSY contributions to
the $b\rightarrow s\gamma$ process, both in the minimal supersymmetric
standard model (MSSM) and in the minimal
supergravity model, we re-evaluate the branching ratio and present the result
as a constraint on charged Higgs mass from this process\cite{GO}.
Next, we calculate the $B^0_d-\bar{B}^0_d$ mixing parameter ($x_d$)
and the CP violating parameter in the $K^0-\bar{K}^0$ mixing (
$\epsilon _K$) in the SUSY model.
Here also there have been many previous works\cite{CPsusy,BBMR,BBbar}.
We have updated the analysis \cite{GNO} using the new experimental
input on the top quark mass\cite{top} and the constraint from
the measured inclusive $b\rightarrow s\gamma$ branching ratio\cite{cleo}.
Presently,
these processes do not put strong constraints on the SUSY parameter space
once other phenomenological constraints
are taken into account. However, when
the measurements of CP asymmetries at future B factory experiments
provide new information on the CKM matrix
there is a good chance that some SUSY effects are extracted from combined
analysis of  $x_d$, $\epsilon _K$ and the CP asymmetries in B decays.

\vglue 0.6cm
{\elevenbf\noindent 2. Minimal Supergravity Model}
\vglue 0.4cm
We consider here the minimal supersymmetric standard model (MSSM)
\cite{Nilles}. Each
particle of the SM has its superpartner. Corresponding to gauge bosons, quarks,
leptons and Higgs boson, gauginos, squarks, sleptons and
higgsino have to be introduced.
These particles can be classified in five sectors,
namely, three ordinary sectors (quark-lepton, gauge field and Higgs
sectors) and two sectors of superpartners (squark-slepton and
chargino-neutralino-gluino sectors). Here the charginos and neutralinos
are combinations of the gauginos and higgsinos.

The Higgs sector contains two Higgs doublets. One of them, denoted by
$H_1$, couples to the down-type quarks and the leptons and the other,
denoted by
$H_2$, to the up-type quarks. There are five physical particle states, i.e.,
two neutral scalars and one neutral pseudo scalar and one pair of
charged Higgs. The masses and mixings of these particles
are parametrized by the pseudo scalar
mass ($m_A$) and the ratio of the two vacuum expectation values,
$\tan {\beta}=\frac{<H^0_2>}{<H^0_1>}$, where $H^0_1$ and  $H^0_2$
are neural components of the Higgs fields. In addition to these
parameters the top and stop masses enter in the formulas of the
various Higgs masses through one loop corrections to the Higgs
potential\cite{SUSYHiggs}.

For discussions on FCNC processes the existence of the
charged Higgs is important. At the tree level, the charged Higgs
and the pseudo scalar masses are related as follows;
\begin{equation}
m_{H^\pm}^2=m_W^2+m_A^2,
\end{equation}
where  $m_{H^\pm}$ and $m_W$ are masses of the charged Higgs
and the W boson. This relation remains to be in a good approximation
even if the one loop corrections to the Higgs potential are taken
into account\cite{charged}. Notice that properties of the lightest neutral
scalar Higgs are almost the same as those of the SM Higgs
if the charged Higgs mass is much larger than 200 GeV. On the other
hand, the investigation of the lightest Higgs
alone could provide us an evidence that the Higgs sector is different from
the simplest one Higgs doublet model in the case that the
charged Higgs is relatively light. We will see later that
the FCNC and CP violating processes are especially sensitive to the
parameter region with the light charged Higgs.

The chargino-neutralino sector consists of two charged
Dirac fermions and four neutral Majorana fermions. Their mass matrices
depend on SUSY breaking gaugino masses ($M_1, M_2$) and a higgsino
mass parameter ($\mu$) as well as the W and Z boson masses and
$\tan{\beta}$. The chargino mass matrix is given by
\begin{eqnarray}
M_C &=&  \left( \begin{array}{cc}
                M_2 & \sqrt{2}m_W \sin \beta\\
                \sqrt{2}m_W \cos \beta & \mu\\
                \end{array} \right),
\end{eqnarray}
where the first row and column corresponds to the wino (a superpartner
to the W gauge boson) states and
the second row and column to the higgsino states. From this mass
matrix we can see that the lightest chargino  behaves like a pure
wino (or higgsino) in the limit of $M_2\ll \mu$ ($M_2\gg \mu$ ).

In the squark sector one complex field must be introduced for each Weyl fermion
in the quark sector. Since there are three generations of quarks
the up-type and down-type squark mass matrices are 6x6
matrices including left-right squark mixing terms. These squark
mass matrices depend on SUSY soft breaking parameters. A general
form in the Lagrangian contributing to the matrices is given by
\begin{eqnarray}
{\cal L}_{soft} &=& -m_{Q_{ij}}^2\tilde{q}_{L_i}^*\tilde{q}_{L_j}
                  -m_{d_{ij}}^2\tilde{d}_{R_i}^*\tilde{d}_{R_j}
                  -m_{u_{ij}}^2\tilde{u}_{R_i}^*\tilde{u}_{R_j}
\nonumber\\
              & & -\tilde{u}_{R_i}^*(m_u A_u)_{ij}\tilde{u}_{L_j}
                  -\tilde{d}_{R_i}^*(m_d A_d)_{ij}\tilde{d}_{L_j}
	          + c.c..
\end {eqnarray}
Although these soft breaking terms contain many free parameters
in general SUSY SM, their forms are strongly constrained
from FCNC processes. Especially, the smallness of the $K^0-\bar{K}^0$
mixing requires that the masses of squarks with the same gauge quantum
numbers should be highly degenerate for the first and second generations
if these masses are less than a few TeV\cite{KKbar}. Whether this degeneracy is
obtained without fine-tuning depends on how these SUSY soft breaking
terms are generated.

In the minimal supergravity model, soft breaking terms are supposed
to arise from gravity interactions with a sector where local SUSY is
spontaneously broken. We can assume that this sector, called
a hidden sector, is flavor-blind. In such a case the ordinary Yukawa
coupling constants in the superpotential are a unique source of
the flavor mixing. In this minimal model the soft
breaking terms are given as,

\begin{equation}
{\cal L}_{soft} = -m_0^2 \sum_{i}|\phi_i|^2
                    -A (H_2 \tilde{u}^c f_u \tilde{q}_L+
                        H_1 \tilde{d}^c f_d \tilde{q}_L) +c.c.,
\end{equation}
where $m_0^2$ is a common SUSY breaking mass for all scalar fields and
$A$ is a common trilinear coupling parameter. $f_u$ and $f_d$ are the
ordinary Yukawa coupling constant matrices.

These parameters are supposed to be generated at the GUT or Planck
scale. The squark mass matrices at the weak scale are calculated by
solving renormalization group equations for soft breaking and other
relevant parameters. As a result, the following general conclusions
can be drawn:\\

\noindent 1) The squarks in the first and second generations with the same
quantum numbers remain highly degenerate at the weak scale.
Therefore, the constraint from the $K^0-\bar{K}^0$ mixing is
naturally satisfied. On the other hand, the squarks in the third
generation can be substantially lighter than other squarks
because renormalization effects due to the large top Yukawa
coupling constant make the stop or left-sbottom mass lighter
at the low energy scale\cite{KKSG}.\\

\noindent 2) The electroweak symmetry breaking can be induced by the
renormalization effects due to the large top Yukawa coupling
constant starting from the assumption that all the scalar fields have
a common SUSY breaking mass.  This is called radiative electroweak symmetry
breaking scenario\cite{RBS}.\\

\noindent 3) Although squarks and quarks are simultaneously diagonalized
at the GUT scale, the renormalization from the GUT to the weak scale can induce
a mismatch between two mass matrices. This will cause flavor changing
interactions even in gluino(or neutralino)-quark-squark couplings
\cite{KKSG}.\\

We have solved the renormalization group equations numerically
including full complex Yukawa coupling matrices and calculated the squark
mass matrices at the weak scale. In the present model, the squark's
flavor mixing is completely determined by the CKM matrix
and other flavor-blind parameters. We have also found
that in a very good approximation the complex phase of the squark's
mixing matrix is the same as
that of the corresponding element in the CKM matrix. As a result, the
$B^0-\bar{B}^0$ box diagrams both for the SM contributions and for the SUSY
ones have the same complex phase.
Previously, this was pointed out
using an approximate solution of the renormalization group equations
\cite{CPsusy}.
Our numerical calculation has confirmed the assertion.

Although the  $K^0-\bar{K}^0$ mixing does not cause any serious
problem once we take the minimal flavor mixing scenario, the FCNC
processes including the third generation might be substantially
influenced by the presence of SUSY particles. We will discuss such
processes, i.e. $b\rightarrow s\gamma$, $B^0_d-\bar{B}^0_d$ mixing,
and $\epsilon _K$ in the followings.

\vglue 0.5cm
{\elevenbf \noindent 3. $b\rightarrow s\gamma$ Process in the Minimal
Supergravity Model}
\vglue 0.4cm

Recently, the CLEO collaboration reported the inclusive branching ratio
of the radiative b decay, $Br(b\rightarrow s\gamma)=(2.32\pm 0.51 \pm
0.29 \pm 0.32)\times 10^{-4}$\cite{cleo}. This value is consistent with the
theoretical prediction within the SM which is $(2\sim 3)\times 10^{-4}$
\cite{GSW,Misiak,BurasMisiakMunzPokorski}.
In the SM this process is induced by the electroweak one-loop diagram
called a penguin diagram. If new physics beyond the SM exists, there may
be extra contributions to the $b\rightarrow s\gamma$ amplitude.
In fact, for a certain type of two Higgs doublet model (THDM) called Model II
the charged Higgs mass less than 260 GeV is excluded by this process
\cite{cleo,GSW,thdm}.
In SUSY models, SUSY particles also contribute to the $b\rightarrow s\gamma$
amplitude in addition to the SM particles and the charged Higgs.
New contributions are loop diagrams from (i) chargino and
up-type squarks, (ii) gluino and down-type squarks and (iii)
neutralino and down-type squarks. Although the charged Higgs contribution
is the same as that of the Model II THDM, and therefore only
constructively interferes with the SM contribution, the amplitude due
to the SUSY particle loops can have either sign depending on parameters
in the SUSY Lagrangian. Therefore, no general bound on the charged Higgs
mass is obtained unlike the simple Model II THDM.

We now consider the $b\rightarrow s\gamma$ branching ratio in the
context of the minimal supergravity model and present the results
as the charged Higgs mass bounds\cite{GO}.
In this case number of free parameters in the SUSY sector is much
smaller than that in the general SUSY SM. Requiring
the radiative electroweak symmetry breaking, we can take $\mu, M_2,
\tan{\beta}$ and the charged Higgs mass as independent parameters
after using the GUT relations among SU(3), SU(2) and U(1) gaugino masses.
Then, all other Higgses' and squarks' masses and mixing parameters
can be calculated with the help of the renormalization group equations.
The detail of our calculation is described in Ref. \cite{GO}.

In Fig. 1 the $b\rightarrow s\gamma$ branching ratio is shown for
$m_t$ (top mass) $= 175$  GeV and $\tan{\beta}=5$. Each point of this
figure corresponds to a particular choice of free parameters.
\begin{figure}
\baselineskip=12pt
	\vspace{8cm}
	\caption{\tenrm{$b\rightarrow s\gamma$ branching ratio for
$m_t= 175$  GeV and $\tan{\beta}=5$. Each dot corresponds to a sample
point which satisfies radiative breaking and phenomenological constraints.
Solid line represents the branching ratio calculated with the SM and
charged Higgs contributions only (Model II THDM). Dot-dashed line
represents the SM value.
}}
\end{figure}
In the calculation we have taken account of various phenomenological
constraints \cite{pdg}:
(i) the mass of any charged SUSY particle is larger than 45 GeV,
(ii) the sneutrino mass is larger than 41 GeV,
(iii) the gluino mass is larger than 100 GeV,
(iv) neutralino search results at LEP \cite{aleph}, which require
$\Gamma(Z \rightarrow \chi \chi)< 22$ MeV,
$\Gamma(Z \rightarrow \chi \chi')$,
$\Gamma(Z \rightarrow \chi' \chi')< 5 \times 10^{-5}$ GeV,
where $\chi$ is the lightest neutralino and
$\chi'$ is any neutralino other than the lightest one,
(v) the lightest SUSY particle (LSP) is neutral,
(vi) the condition for not having a charge or color
  symmetry breaking vacuum \cite{aterm}.
We have neglected the neutralino loop contribution to the amplitude
which is known to be very small. For comparison we also present
the branching ratio which is calculated with only the SM and charged Higgs
contributions retained. We can see that the predictions of the branching
ratio are divided by the line of the THDM. In fact points above the line
corresponds to the case $\mu < 0$ and below it to the case $\mu > 0$,
respectively. Therefore, whether the SUSY particle effects enhance
or suppress the branching ratio depends on the sign
of the $\mu$ parameter, which was pointed out previously\cite{Lopez1,
BertoliniVissani,Lopez2}.
In Figs. 2 and 3,
excluded region in the charged Higgs mass and $\tan{\beta}$ space
from the $b\rightarrow s\gamma$ process is shown separately for $\mu < 0$
and $\mu > 0$.
\begin{figure}
\baselineskip=12pt
	\vspace{8cm}
	\caption{\tenrm{Excluded region in the $\tan{\beta}$ and
$m_{H^\pm}$ space for $\mu < 0$. Each line represents the lower
bound for the charged Higgs mass: solid line: all constraints included;
dashed line: without $b\rightarrow s\gamma$ constraint; dot-dashed line:
Model II THDM with  $b\rightarrow s\gamma$ constraint.
}}
\end{figure}
\begin{figure}
\baselineskip=12pt
	\vspace{8cm}
	\caption{\tenrm{The same as Fig.2 for$\mu > 0$ }}
\end{figure}
In accordance with the above discussion, the lower bound of the charged
Higgs mass becomes much larger than that in the Model II THDM for
$\mu < 0$, but no strong bound is obtained for  $\mu > 0$ due to
cancellation between
the SUSY and other contributions. In determining the excluded region
in the parameter space, we have calculated the $b\rightarrow s\gamma$
branching ratio varying free parameters ($\mu, M_2$) for each fixed
set of the charged Higgs mass and $\tan{\beta}$. Since the main
theoretical ambiguity comes from the choice of the renormalization
scale ($Q$) at the bottom scale in evaluating the QCD correction
\cite{BurasMisiakMunzPokorski}, we have
calculated the branching ratio by varying the renormalization scale
from $Q=m_b/2$ to $Q=2m_b$ where the bottom mass $m_b$ is taken to be
4.25 GeV. To be conservative, we also include additional 10 \%
theoretical uncertainties. Then, if the calculated branching ratio
cannot be within the experimental value ($1\times 10^{-4}<Br<
4\times 10^{-4}$) for any choice of ($\mu, M_2$) even if the theoretical
uncertainties are taken into account, we regard the point in the charged
Higgs mass and $\tan{\beta}$ space excluded. In Fig. 4, two cases,
$\mu < 0$ and $\mu > 0$ are combined and the excluded region is
shown independently of the sign of $\mu$. For $3\lsim \tan{\beta} \lsim
5$, the charged Higgs mass smaller than 180 GeV is excluded by this
process.
\begin{figure}
\baselineskip=12pt
	\vspace{8cm}
	\caption{\tenrm{Excluded region in the $\tan{\beta}$ and
$m_{H^\pm}$ space irrespective of the sign of $\mu$. The meaning of
the lines are the same as those in Fig. 2.  }}
\end{figure}
Due to the phenomenological constraints and the condition for radiative
electroweak symmetry breaking,
this region was previously allowed only for $\mu < 0$.
It is, however, completely excluded by the new
measurement of the $b\rightarrow s\gamma$ process.
\vglue 0.5cm
{\elevenbf \noindent 4. $x_d$, $\epsilon _K$ and CP asymmetries
in B decays \hfil}
\vglue 0.4cm
In the SM, all the flavor mixing and CP violating processes are determined
by the parameters of the CKM matrix. There are four physical parameters
in this matrix. A convenient way of the parametrization was given by
Wolfenstein as follows\cite{Wolf}:
\begin{eqnarray}
V_{CKM}&\simeq&  \left( \begin{array}{ccc}
                1-\frac{\lambda^2}{2}& \lambda
                 & \lambda^3 A (\rho-i\eta)\\
                -\lambda  & 1-\frac{\lambda^2}{2} & \lambda^2 A\\
                \lambda^3 A (1-\rho-i\eta)& -\lambda^2 A & 1\\
                \end{array} \right),
\end{eqnarray}
where we have ignored higher order terms in $\lambda$ in each
element. Among four parameters the Cabbibo mixing parameter $\lambda$
and the $A$ parameter which is determined by $|V_{cb}|$ are
well known experimentally. The $\rho$ and $\eta$ parameters
are not yet precisely fixed. We can determine the present allowed
region on the ($\rho$, $\eta$) space from three measured quantities,
i.e., the CP violation parameter in K decays ($\epsilon _K$),
the $B^0_d-\bar{B}^0_d$ mixing parameter ($x_d$) and $|V_{ub}|/|V_{cb}|$
from charmless b decays.

When experiments on B decays are done at the asymmetric B factories
we can directly measure three angle of the triangle formed by
three points $0+i0$, $1+i0$, $\rho+i\eta$ in the complex plane. This is called
the unitarily triangle. The time dependent asymmetry of the $B^0_d$
($\bar{B}^0_d$)
decay to a CP eigenstate ($f$) determines a quantity $\xi=Im \left(
\frac{q}{p}\right)_B \left(\frac{\bar{A}_f}{A_f}\right)$, where
$\left(\frac{q}{p}\right)_B$ corresponds to the complex phase
of the $B^0_d-\bar{B}^0_d$
box diagram and $A_f(\bar{A}_f)$ is the amplitude of $B^0
\rightarrow f (\bar{B}^0\rightarrow f)$\cite{Sanda}. In the SM the $B^0
\rightarrow \psi K_s$ mode gives the angle $\phi_1$ ($=$ angle
at the point $1+i0$) which is in fact directly related to the phase
of the $B^0_d-\bar{B}^0_d$ box diagram. $\phi_2$ ($=$ angle
at the point $\rho+i\eta$) is obtained from $B^0
\rightarrow \pi \pi$ or $\pi \rho$ process. The measurement of the
angle $\phi_3$ ($=$ angle at the point $0+i0$) can be done with direct CP
violation in $B \rightarrow D K$ processes\cite{Gronau}. Combining these
measurements we will be able to determine the $\rho$ and $\eta$
parameters precisely. If the SM prediction is correct the parameter
determined by these angle measurements should fall in the range
obtained by the analysis of $\epsilon _K$,
$x_d$ and $|V_{ub}|/|V_{cb}|$. Any inconsistency among these
observables give us a hint on physics beyond the SM.

In the minimal supergravity model the phase of the $B^0_d-\bar{B}^0_d$
box diagram is the same as that of the SM box diagram in a very
good approximation, as we discussed in section 2. On the other hand
the magnitude of the mixing can be different from the SM due to
contributions from the charged Higgs and SUSY particles.
The same is true for $\epsilon _K$ which is determined from the
$K^0-\bar{K}^0$ box diagram. Therefore, combining these observables
with angle informations we may be able to get useful information
on existence of SUSY particles.

We present calculation of  $x_d$ and $\epsilon _K$ in the minimal
supergravity model\cite{GNO}. In addition to the SM box diagram, we have
included diagrams with the charged Higgs, chargino and up-type
squarks, gluino and down-type squarks. One loop diagrams
with neutralino are expected to be very small, therefore neglected
here. In the calculation of these quantities we have taken into account
the phenomenological constraints and the condition for the radiative
electroweak symmetry breaking as described in the previous section.
We have also included the constraint on the SUSY parameter space from the
$b\rightarrow s\gamma$ process discussed in section 3. Fig. 5 shows
a plot of $x_d=\frac{\Delta M_B}{\Gamma}$ in the SUSY model normalized
by $x_d$ in the SM.
\begin{figure}
\baselineskip=12pt
	\vspace{8cm}
	\caption{\tenrm{Ratio of $x_d$ in the supergravity model
and that of the SM for $m_t= 175$  GeV and $\tan{\beta}=2.5$.
Each dot correspond to a sample point
which satisfies radiative breaking and phenomenological constraints
including the \bsg constraint.
Solid line represents the same ratio calculated with the SM and
charged Higgs contributions only (Model II THDM). }}
\end{figure}
This ratio does not contain the theoretical ambiguity of the hadron
matrix element, i.e., $f_B^2 B_B$. Here, we have taken $m_t=175$ GeV,
$\tan{\beta}=2.5$ and $\rho=0.18,\eta=0.31$. In the supergravity
model this ratio is almost independent of the values of $\rho$ and $\eta$
since the diagram containing squarks has almost the same
$\rho$ and $\eta$ dependence as the SM box diagram.
Here we have also shown the line
corresponding to the case in which only the SM and the charged Higgs
contributions are retained. We can see that the SUSY contributions
always enhance the value of $x_d$.
This is in contrast to the $b\rightarrow s\gamma$ case where
cancellation between SUSY and other contributions is possible.
(This is true even if we have not included the $b\rightarrow s\gamma$
constraint.) The SUSY and the charged Higgs contribution typically
enhance $x_d$ by 10 $\sim$ 20 \%. The enhancement is larger
for smaller value of the charged Higgs mass. We also calculated
this ratio for different values of $\tan{\beta}$. The result
show that $x_d$ is larger for smaller $\tan{\beta}$. A similar
plot is shown for $\epsilon_K$ in Fig. 6. The size of the SUSY
and charged Higgs contributions is quite similar to that of $x_d$.
\begin{figure}
\baselineskip=12pt
	\vspace{8cm}
	\caption{\tenrm{The same figure as Fig.5 for the ratio
of $\epsilon_K$ in the minimal supergravity model and that of SM.
The solid line corresponds to Model II THDM.}}
\end{figure}

Since the effect of new particles are less than 20 \%, $x_d$ and
$\epsilon_K$ do not strongly constrain the SUSY parameter space from
the present measurements. This is because large
hadronic uncertainties exist in the calculation
of $\epsilon_K$ and $x_d$, i.e. $B_K$,
and $f_B^2 B_B$, and also still two free parameters $\rho$
and $\eta$ remain in the prediction of the size of $\epsilon_K$
and $x_d$.
The situation will be changed if independent information on the
$\rho$ and $\eta$ parameters is obtained at the B factories. With
expected improvement on determination of $B_K,f_B$ and $B_B$ from
the lattice gauge theory in the next few years\cite{Soni},
the 10 $\sim$ 20 \%
effects on $x_d$ and $\epsilon_K$ can be large enough to detect
new physics contributions like SUSY particles' ones.
\vglue 0.5cm
{\elevenbf \noindent 5. Conclusions \hfil}
\vglue 0.4cm
We have seen that SUSY particles and charged Higgs can contribute to
various FCNC and CP violation processes like $b\rightarrow s\gamma$,
$B^0-\bar{B}^0$ mixing and $\epsilon_K$. They are induced not only
by the usual quark flavor mixing but also by the squark's counterpart.
In the minimal supergravity model this squark's flavor mixing is specified
by a few free parameters so that the model has predictive power. We
have seen that the prediction of the $b\rightarrow s\gamma$
branching ratio depends on the sign of the $\mu$ parameter. Then,
the sensitivity to the charged Higgs mass is quite different
for $\mu>0$ and $\mu<0$. We also
calculated $x_d$ and $\epsilon_K$ in this model and have shown that
these quantities are always enhanced compared to the SM prediction.
The effects can be as large as 20 \% after taking account of the
$b\rightarrow s\gamma$ constraint. These effects are large enough
to give impacts on the new physics search at the B factories by
measuring sides and angles of the unitarily triangle.
\vglue 0.5cm
{\elevenbf\noindent References \hfil}
\vglue 0.4cm

\vglue 0.5cm
\end{document}